\documentclass[12pt]{article}

 \input epsf

\input psfig

\newcommand{\be}{\begin{equation}}
 \newcommand{\ee}{\end{equation}}

\begin{document} 

\begin{center}
{\large {\bf Neutrino Zero Modes and Stability of Electroweak Strings}}

\vspace{1cm}
Dejan Stojkovic 
\vspace{.5cm}

\it
Department of Physics \\
Case Western Reserve University \\
Cleveland, OH 44106-7079, USA \\
\vspace*{0.5cm}

\end{center}


\begin{abstract} 
We discuss massless and massive neutrino zero modes in the
background of an electroweak string.  We argue that the eventual absence of the
neutrino zero mode implies the existence of topologically stable strings where
the required non-trivial topology has been induced by the fermionic sector.
\end{abstract}

\vspace{.5cm}

Recent experimental evidence strongly suggests that neutrinos are very light,
but not massless.  A simple way to extend the standard model such that it
incorporates light neutrino masses is to add a singlet right-handed neutrino.
The origin of $M_R$ is a vacuum expectation value of some scalar field, acquired
at some energy scale much greater than that of the electroweak scale.  $M_R$ can
therefore be taken to be spatially homogeneous.  The Lagrangian is:

\begin{eqnarray} \label{LSM} 
{\cal L}_{SM} &=&
-{1\over4}W^a_{\mu\nu}W^{a\mu\nu}-{1\over4} F_{\mu\nu}F^{\mu\nu}+
\left(D_\mu\Phi\right)^\dagger\left(D^\mu\Phi\right) - 
\lambda\left(\Phi^\dagger\Phi - \eta^2\right)^2 \nonumber\\ &-&
i{\bar\Psi}\gamma^\mu D_\mu\Psi - i\overline{e_R}\gamma^\mu D_\mu e_R + 
h'\left(\overline{e_R}\Phi^\dagger\Psi+ {\bar\Psi}\Phi e_R\right) \\ &+& h
\left[\bar\Psi i \tau_2 \Phi^\ast \nu_R + \overline{\nu_R}\Phi^T 
(i\tau_2)^{\dagger} \Psi\right] + {1\over2} \overline{\nu_R^c} M^\ast_R \nu_R +
{1\over2} \overline{\nu_R} M_R\nu_R^c \nonumber  
\end{eqnarray} 
with
$\Psi\equiv(\nu_L, e^-_L)^T$, $\Phi \equiv (\phi^+,\phi)^T$, 
$\nu^T_L =(\alpha ,\beta , -\alpha , -\beta )$, 
$\nu^T_R =(\gamma ,\delta , \gamma , \delta )$,
$\nu^c \equiv C \bar{\nu}^T$ and with the usual definition of the gauge and
Higgs fields and covariant derivatives \cite{nzms}.  In the Z-string ansatz, all the gauge
and Higgs fields are zero except:  
\be 
Z^\mu \sim \left(0,-{{v(r)}\over r}{\vec
e}_\theta\right), \quad \phi=\eta f(r)e^{i\theta} \label{zstringfields}
\ee
where $v(r)$ and $f(r)$ are the Z-string profile functions.  The neutrino 
equations of motion are:

\begin{eqnarray} 
i\gamma^\mu D_\mu \nu_L &=& h\phi^\ast \nu_R \nonumber\\
i\gamma^\mu \partial_\mu \nu_R &=& h\phi \nu_L + M_R (\nu_R)^c
\label{Diraceqns}
 \end{eqnarray}

Using the ansatz for $\beta$, $\beta = \sum_{m=-\infty}^{\infty} i \beta_m(r)
e^{ik_z z -i\omega t + i m\theta}$ and similarly for the other spinor
components \cite{nzms}, after setting $\omega = k_z = 0$, we get a set of recursive
equations for the coefficients $\beta$ and $\gamma$:

\begin{eqnarray} \label{sys} \beta'_m + {(m+v)\over r}\beta_m & = &-h\eta f
\gamma_{m} \nonumber \\ \gamma'_m - {m\over r} \gamma_m - iM_R
\gamma^\ast_{-1-m} &=& -h\eta f \beta_{m} \end{eqnarray}

An analytic solution to the system (\ref{sys}) could not be found, but we can
learn something about the structure of solutions by looking at their asymptotic
behavior.  For $m = 0$, at large r, where $f \sim 1$, $v \sim 1$ the system
(\ref{sys}) has four solutions for each spinor component.  Two of them are
exponentially growing while two are exponentially decaying.  Only the latter
ones are physically acceptable.  For small $r$ where, $v \sim v_0 r^2$ and $f
\sim f_0 r$, the structure of the system (\ref{sys}) is such that only three
solutions are well behaved i.e.  nonsingular and singe valued.  With three
well-behaved solutions at the origin and two well-behaved solutions at infinity
there must be at least one solution which is well-behaved everywhere.  Each of
the two well-behaved solutions at infinity matches on to a unique linear
combination of all four solutions at the origin where only one of them is bad,
so there is always one linear combination of the two good solutions at infinity
which does not have any contribution from the single bad solution at the origin.
We conclude that in this case there is always one well behaved zero mode
solution.

In the case of massless neutrino, for large $r$, $\omega \neq 0$ and $m = 0$,
we get Bessel's equation for $\beta$, 
\be
\beta''_0+\frac{1}{r} \beta'_0+\beta_0 (\omega^2 - \frac{1}{r^2})=0 
\ee 
with the general solution given by a linear combination of Bessel's functions:  
\be
\label{wave} \beta_0(\omega r) = A \sqrt{\omega} J_1(\omega r) + B \sqrt{\omega}
Y_1(\omega r) 
\ee
where $A$ and $B$ are constants.
We see, therefore, that the solution is properly
$\delta$-function normalized.  In the zero mode limit $\omega \rightarrow 0$.
Using the asymptotic expansion for Bessel's function of small argument we get
$\beta_0 \sim 1/r$ which is the result found in most of the literature.  The
usual interpretation is that this state is not normalizable.  If we look at the
solution for $\beta_0 $ as an isolated zero mode, it is obviously not
normalizable because the normalization integral diverges logarithmically 
for large $r$.  But,
we have seen that this state is actually part of a continuum spectrum of the
theory and is Dirac delta function normalizable.  Therefore it is a valid zero
mode solution.  (A similar analysis can be done for zero modes of massive and
massless fermions in the background of a domain wall \cite{nzmdw}.)

In the most general case \cite{nmm,gc}, the presence of a constant left-handed
Majorana mass, $M_L$, could destroy the neutrino zero mode. The left-handed
Majorana mass terms have the form:
\be
\overline{\nu_L} M_L\nu_L^c + \overline{\nu_L^c} M^\ast_L \nu_L \ .
\label{mlterms}
\ee
It was shown in \cite{gc,tibf} that for a certain range of mass parameters the 
neutrino zero mode does not exist on a electroweak string.  We will argue that
this fact imply the existence of the phenomenon which we call ``topology
induced by fermions".

\begin{figure}[ht]
\centerline{\epsfxsize = 0.50 \hsize \epsfbox{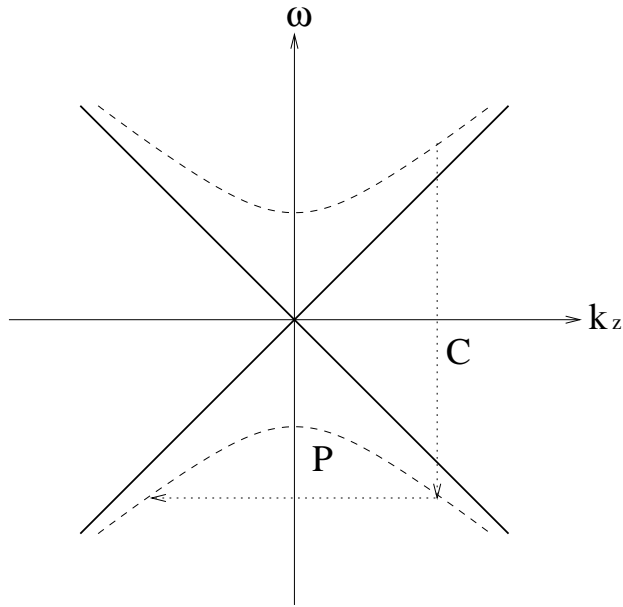}} 
\caption{\label{tzm}
In the usual case, the fermionic spectrum contains two lines (solid)
that describe zero modes boosted in the $z$-direction. One line
describes the left-moving and the other the right-moving zero modes. 
Perturbations of the bosonic background can mix 
the two zero modes at the origin and convert them into two massive
modes that lie on hyperbolae. The points on the two hyperbolae
are related by CP transformations as shown.}
\end{figure}

\begin{figure}[ht]
\centerline{\epsfxsize = 0.50 \hsize \epsfbox{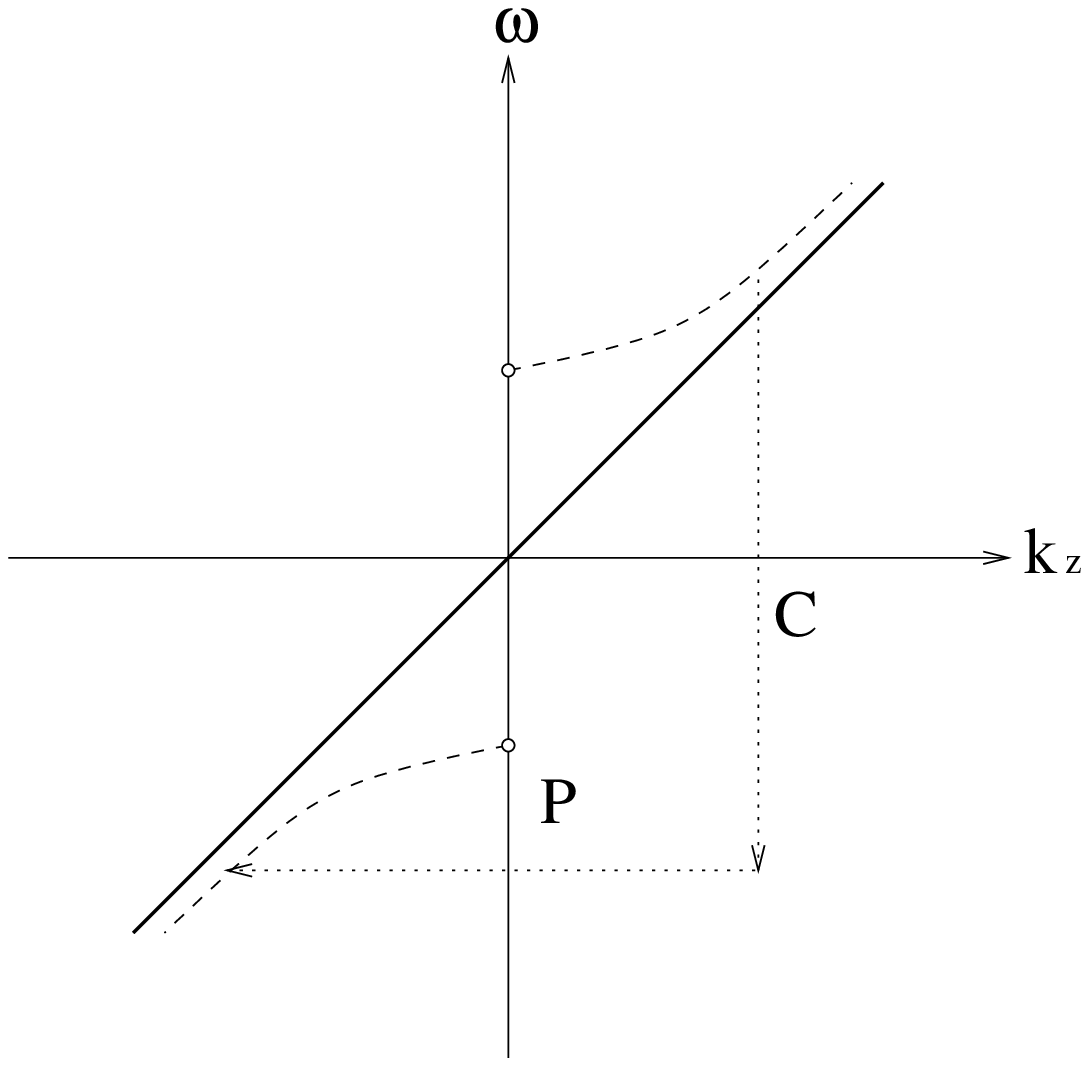}} 
\caption{ \label{ozm}
In the present case, the spectrum contains only the line of boosted 
electron zero modes (the $\omega = +k_z$ line). Perturbations of
the background cannot change the line into the spectrum for massive 
fermions (hyperbolae). In particular,
perturbations cannot convert the single zero mode ($\omega =0=k_z$)
into a massive mode because CP invariance would require it to
split into two modes, leading to a discontinuous spectrum. }
\end{figure}

The possibility of not having a neutrino zero mode while the
corresponding electron zero mode is still present is important from the point of
view of the string stability under perturbations.  In the usual case with a pair
of zero modes (see Fig. (\ref{tzm})), perturbations in the Higgs and gauge sectors mix the two zero
modes and convert them into two low lying massive states.  If the neutrino zero
mode is not present, perturbations of the string cannot lift the electron zero
mode into a massive mode without violating CP invariance (see Fig. (\ref{ozm}))
(the bosonic background
is CP invariant since we can absorb all the CP violating  phases into the spinor
components). This conclusion that a single zero mode can not be removed from
the spectrum, obtained from pure symmetry consideration, 
can be confirmed by studying the stability of fermionic zero modes on
electroweak strings under perturbations \cite{thesis}. For non-topological 
strings this conclusion seems paradoxical
because there is no topology in the bosonic sector that prevents the string from
decaying into the vacuum.  However, there is no electron zero mode in the
vacuum.  The only resolution seems to be that the absence of the neutrino zero
mode provides topological stability to the string and that the topology enters
via the fermionic sector of the model.

It is difficult to incorporate a constant $M_L$ into the standard
model as a vacuum state.  The source of $M_L$ could be the vacuum expectation
value of an $SU(2)_L$ Higgs triplet $\Delta$.  The coupling of $\Delta$ to the
lepton doublet is given by $\overline{\Psi^c} i\tau_2 \Delta \Psi$ .  The
kinetic terms in the Lagrangian now contain the covariant gradient of the Higgs
triplet which couples $\Delta$ to the Z-gauge field.  This coupling forces a
solution of the bosonic equations of motion for $\Delta$ to wind around the
Z-string and hence to vanish on the string axis.  Similarly, a cubic interaction
term $\Phi^\dagger i \tau_2 \Delta \Phi^\ast$ forces the triplet to wind due to
$\theta$ dependence of $\Phi$ .  The way to avoid the problem of coupling to the
gauge fields is to set all the standard model gauge fields to zero.  In this
case the gauge groups become global.  We can also set a cubic interaction term
$\Phi^\dagger i \tau_2 \Delta \Phi^\ast$ to zero.  This can be done by requiring
the lepton number conservation in the initial Lagrangian.  If we consider the
situation in which the doublet and the triplet get vacuum expectation values:
\be 
\Phi = \pmatrix{0\cr \eta f(r)e^{i\theta}}\ , \ \ \Delta =
\pmatrix{0 & 0 \cr M_L & 0} \label{background} 
\ee 
the symmetry
breaking pattern is $SU(2)_L \times U(1)_Y \rightarrow U(1)_Q$ where all the
symmetry groups are global.  The model, although not the standard model anymore,
supports a spatially homogeneous $M_L$ and the non-topological global Z-string,
and the above discussion applies to it.

From the other point of view, if we want to keep all the features of the
standard model, at a given moment of time, one can consider
fermions with the general mass matrix in the background of the standard model
Z-string.  We again consider the symmetry breaking pattern (\ref{background}).
As is characteristic for a phase transition phenomenon, the Higgs and gauge fields,
initially take different random configurations at different regions in space.
One of the possible configurations we would like to analyze is a configuration
in which $\phi$ winds around some axis while $M_L$ and $M_R$ are constant.  If a
non-trivial topology was present in the bosonic sector of the model, such a
configuration would be stable.  However the present model does not have any
non-trivial bosonic topology.  A configuration with the constant $M_L$ is not a
solution of the equations of motion and does not minimize the energy.  Usually,
such a non-equilibrium configuration evolves continuously into a state of
minimum energy which, in this case, is the vacuum where all the fields are constant,
unless there are topological sectors which forbid the evolution from one sector
to another.  The bosonic sector of the standard model does not contain such 
topological sectors.  However, if the parameters of the model are such that there is no
neutrino zero mode \cite{gc,tibf}, then the constant $M_L$ configuration
supports a single electron zero mode.  As we have argued, CP invariance forbids
removing (or creating) a single zero mode and different configurations can not
continuously transform into one another.

Although we did not show that  there is a topological charge density 
which can be associated with this ``fermionic topology", we showed that 
there is an  integer $n_L - n_R$ (the
difference between the left- and right- moving fermionic zero modes, namely
electron and neutrino zero modes) that characterizes the field configuration.
We argue that $n_L - n_R$ plays the role of the topological charge and cannot
change under continuous evolution of the fields.  Since $n_L-n_R$ in the vacuum
is zero while in our chosen background is 1, the background and the vacuum are in
distinct topological sectors.
Thus, the configuration with a single electron zero mode leads to a stable string
where the non-trivial topology arises from the fermionic sector of the theory
and not, as is usual, from the non-trivial topology associated with the bosonic
vacuum manifold.

A real challenge would be a construction of an index theorem describing
this phenomenon. In this case, there should exist a topological charge density 
associated with this ``fermionic topology''. Unlike the all known
versions of (topological) index theorems which take  only the bosonic field
configurations into account, this new index theorem must also depend on 
the fermionic sector of the theory.

\end{document}